\documentstyle[pre,aps,epsf,twocolumn]{revtex}

\begin{document}
\draft
\title{Modelling of spatio-temporal patterns in bacterial 
colonies}
\author{A.M. Lacasta, I.R. Cantalapiedra, C.E. Auguet,
A. Pe\~naranda and  L.\ Ram\'{\i}rez-Piscina }
\address{
Departament de F\'{\i}sica Aplicada,\\
Universitat Polit\`{e}cnica de Catalunya,\\
Avda. Dr. Mara\~{n}on 44, E-08028 Barcelona, SPAIN.}
\date{\today}
\maketitle

\begin{abstract}
A diffusion-reaction model for the growth of 
bacterial colonies is presented.
The often observed cooperative behavior developed by
bacteria which increases their motility
in adverse growth conditions 
is here introduced as a nonlinear diffusion term.
The presence of this mechanism depends on a response
which can present hysteresis. 
By changing only the
concentrations of agar and
initial nutrient, numerical integration of the proposed model
reproduces the different patterns shown by 
{\it Bacillus subtilis} OG-01.

\end{abstract}


\section{Introduction}

Some kinds of bacterial colonies
present interesting
structures during their growth 
\cite{Mat,ben1,ben2,budrene,shapiro,rauprich,williams,rafols,Matsus,Ohg,fujikawa}.
Depending on the bacterial species and the culture conditions,
colonies can exhibit a great diversity of forms.
In general, the complexity of the growth pattern
increases as the environmental conditions become less favorable.
Bacteria respond to
adverse growth conditions by developing sophisticated strategies
and higher micro-level organization
in order to cooperate more efficiently.
Examples of these
strategies are:
the differentiation into longer-motile bacteria, the production of
extracellular
wetting fluid, the
secretion of surfactants which change the surface
tension or the
chemotactic response to chemical agents produced by bacteria
\cite{Mat,ben1,ben2,budrene,shapiro,rauprich,williams,rafols}.
The experiments are usually made in a petri dish,
which contains a
solution of nutrient and agar. A drop of
bacterial solution
is then inoculated  in the center of the dish.
The growth conditions are controlled by
the initial concentration of the medium components.
The agar concentration determines the consistency of the
medium,
which becomes harder as the amount of agar increases, and
the nutrient concentration
controls the bacterial
reproduction.
Depending on these two factors, the colony grows at
a higher or lower
rate, developing different kinds of patterns.

\begin{figure}
\begin{center}
\def\epsfsize#1#2{0.50\textwidth}
\leavevmode
\epsffile{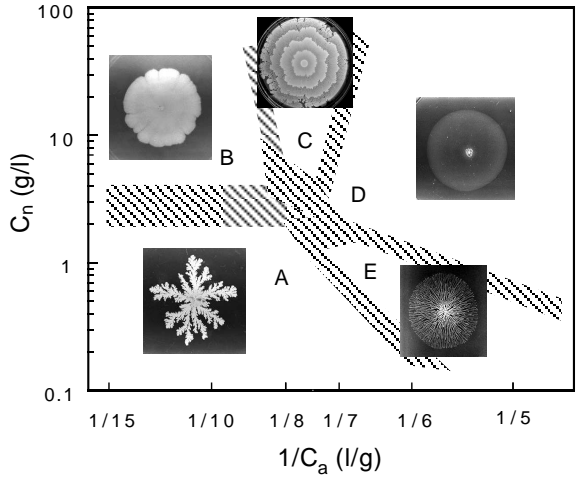}
\end{center}
\caption{
\label{fig1}
Morphological diagram of patterns observed in
colonies of {\it Bacilus subtilis} OG-01
as a function of nutrient concentration ($C_n$) and the
hardness of agar surface ($1/C_a$ being $C_a$ the agar concentration).
Experiments were performed in a petri dish with a diameter of
88 mm. Taken with permission from \protect\cite{rafols}.}
\end{figure}

In particular, colonies of the bacterium
{\it Bacillus subtilis} OG-01 present a rich variety
of structures
\cite{rafols,Matsus,Ohg,fujikawa}.
Fig.\ \ref{fig1} shows the morphological diagram obtained by
Matsushita and co-workers \cite{Ohg}. They classified the
colony patterns into five types, from A to E, whose 
main features can be summarized as
follows.
If the medium is very hard, i.e., with a high
concentration of agar, bacteria can hardly move and
the colony
essentially grows due to the consumption of nutrient
and subsequent
reproduction.
If the level of nutrient is also low (region $A$), the
growth is controlled
by the diffusion of the nutrient
up to the bacteria placed at the interface. The
colony develops a ramified
structure very similar to the patterns obtained with
the diffusion-limited aggregation
model (DLA)\cite{Matsus,witten}. It takes approximately
one month to cover the dish.
If the initial agar concentration remains high and
the nutrient concentration is increased, the growth is
faster than in
region $A$. The branches grow thicker 
until they fuse into a dense disk
with rough interface (region $B$), similar to the
patterns obtained with an
Eden model\cite{vicsek}. This structure needs 5-7
days to cover the disk.
When the level of agar is decreased, which produces a
medium a little
softer than in region B, and the level of
nutrient remains high,
the colony forms concentric rings (region $C$).
This region is characterized by periodic
dynamics:
for 2-3 hours the colony expands while
the bacteria
move actively ("migration phase") and then they
almost stop for 2-3 hours ("consolidation phase"),
during which the
colony does not grow appreciably and the bacterial density
increases due to reproduction.
The crossover between the two phases is sharp.
The periodic cycles of subsequent
migration and consolidation phases create the
pattern of concentric rings \cite{rafols,fujikawa}.
Accurate measurements show that in the growth
phase there is a high concentration of longer
and more motile bacteria, as a consequence of a
differentiation process
\cite{rafols}.
    When a high level of nutrient is maintained, 
and the agar concentration is decreased further,
the colony spreads over the agar plate, and after
less than 8 hours homogeneous disk of  low
bacterial density is
formed covering all the dish (region $D$).
In this thin surface,
bacteria are always short and can move easily by
swimming.
By decreasing the nutrient concentrations for a
semi-solid medium,
the colony develops a densely branched pattern
(region $E$) similar to the
dense branching morphology (DBM) found in other
systems \cite{ben3}.
The ratio of the width of the branches to the gap
between them is constant
over the whole colony. The colony grows quite fast,
showing its main activity at the tips of the fingers, and
covering the dish in less than 24 hours.
The dynamics is related to both the consumption of 
nutrients and the bacterial motility.
In general, when environmental conditions
are adverse (low nutrient or hard surface),
a higher level of cooperation is observed.

The existence of a cooperative behavior seems
to be determinant in the
formation of the rings patterns of region $C$.
The same kind of concentric rings has been
found in experiments with other bacterial
species. In the case of the bacterium {\it Proteus
mirabilis}, the migration phases
clearly involve the movement of 
differentiated swarmer bacteria (elongated
and hyperflagellated)\cite{shapiro,rauprich,williams}.
Similar ring patterns have also been observed in
other non-living systems, like the Liesegang
rings produced by precipitation in the wake of a
moving reaction front \cite{chopard}, or
some experiments of  interfacial electrodeposition
\cite{zeiri}. In the case of the Liesegang
patterns, it is well known that the distance between
rings increases as $t^{1/2}$, whereas in
bacterial and electrodeposition it is
constant.

Several models have been proposed to explain the variety
of patterns exhibited by {\it Bacillus subtilis}, as shown
in Fig.\ \ref{fig1}.
\cite{ben1,ben2,Matsus,Murray,Wakita,kawasaki,mimura,esipov}.
DLA-like patterns (region $A$ in Fig.\ \ref{fig1})
have been interpreted \cite{Matsus} as growth controlled by the diffusion of
nutrients in the context of the DLA model.
Ben-Jacob and co-workers proposed a communicating walkers
model to describe some of the morphologies \cite{ben1,ben2}. This model
reproduces the crossover between regions A and B by coupling random
walkers to fields representing the nutrients. DBM-like patterns are also
obtained by introducing a chemotactic agent.
Other kinds of models are based on reaction-diffusion
equations for bacterial density. The Fisher equation \cite{Murray}
can be used for reproducing the homogeneous circular 
morphology (region $D$) \cite{Wakita}.
Further developments were achieved by introducing new elements to the
Fisher model, such as a field for nutrient
and nonlinear diffusion coefficients \cite{kawasaki,mimura}.
Depending on the new elements
introduced, these models reproduce some of the patterns
of Fig.\ \ref{fig1}.
However, the ring patterns (region $C$) have
so far eluded a satisfactory modelization.
Although the model suggested in Ref.\ \cite{mimura}
can generate concentric ring patterns, they are rather different 
from those observed in experiments \cite{rafols}. 
In fact, dynamical
cycles of consolidation and growth phases are not found. 
Finally, we must mention a model proposed by Esipov {\it et al.}
\cite{esipov} for the study of 
{\it Proteus mirabilis} colonies, which
introduces a life-time for the differentiated swarmer bacteria.
This model reproduces concentric ring patterns but does not explain
why no periodicity is observed in other regions of the morphological
diagram.

In this paper, we propose a model consisting of two
coupled diffusion-reaction
equations for bacteria and nutrient concentrations,
where the bacterial diffusion coefficient can adopt
two different expressions, corresponding to two possible
mechanisms of motion. The first is the usual random
swimming performed by bacteria in liquid medium. The second 
is developed by
bacteria in response to adverse growth conditions,
and depends on their concentration. Bacterial response
is modeled as a global variable that can present hysteresis.
Our model reproduces the five 
morphologies observed
in the experiments, including the ring patterns.

\section{Mathematical Model}
     We consider a two-dimensional system containing
bacteria and nutrients.
Both diffuse, while bacteria
proliferate by feeding on nutrient.
Let us denote by ${\bar b}({\bf r}',\tau)$ the density of
bacteria
at time $\tau$ and spatial position ${\bf r}'$, and by
${\bar n}({\bf r}',\tau)$
the concentration of nutrient. Then, ${\bar b}$ and
${\bar n}$ are in general
governed by the following
equations\cite{Murray,kawasaki}:

\begin{eqnarray}
\frac{\partial {\bar b}}{\partial \tau}&=& \nabla D_b
\nabla {\bar b} +
\theta f({\bar b},{\bar n}),
\nonumber \\
\label{mod1} \\
\frac{\partial {\bar n}}{\partial \tau}&=& D_n \nabla^2
{\bar n} -
f({\bar b},{\bar n}).
\nonumber
\end{eqnarray}

The function $f({\bar b},{\bar n})$ denotes the
consumption term of nutrient by bacteria,
and can be described by Michaelis-Menten
kinetics\cite{Murray}
\begin{equation}
f({\bar b},{\bar n})=\frac{k {\bar n}{\bar b}}{(1+\gamma'
{\bar n})},
\label{growth}
\end{equation}
where $k$ is the intrinsic consumption rate. For small ${\bar n}$,
the consumption rate is approximately linear
in ${\bar n}$ and it saturates at the value
$k/\gamma'$ as ${\bar n}$ increases.
$D_b$ and $D_n$ are the diffusion coefficients of
bacteria and
nutrient respectively. We assume that $D_n$ is
constant, but $D_b$ can depend on
nutrient and bacterium concentrations.

As explained above, experiments show that
in adverse conditions,
bacteria can adapt themselves in order to improve
their motility.
In a soft medium and high nutrient concentration (region $D$ of
Fig.\ \ref{fig1}), short
bacteria can swim randomly without difficulty, but
in an adverse environment (regions $A, B, C$ and
$E$) they need to
develop mechanisms to become more motile.
For intermediate conditions of
semi-solid medium and sufficient nutrient, there are
periods of fast growth
(migration phase) and
slow growth (consolidation phase) that lead to the
concentric ring patterns.

The analysis of periodic rings suggests a dynamical
scheme with hysteresis that can be
outlined in the following way:
during the consolidation phase, the population of
longer-motile bacteria increases in order to
overcome the opposition to the movement. When this
population exceeds a certain
value, enhanced movement becomes possible and
a migration phase begins. Then, however,
a progressive decrease in long-bacterial
population ensues, until it reaches a minimum at
which the "enhanced-movement mechanism" does not
work. Then a new consolidation phase begins.
Within this scheme, region $D$ corresponds to a case
where the maximum value is never
reached (and therefore bacteria always move by
usual diffusion),
whereas in regions $A, B$ and $E$
long-bacterial population does not fall below the
minimum (and therefore always moves
by the enhanced-movement mechanism).
All these ideas can be introduced in our model by
means of two basic points:

(a) The diffusion coefficient $D_b({\bar n},{\bar b})$
can take two
different expressions
depending on the long-bacterial population.

(b) The net production of long bacteria depends on
the environmental conditions and also
on the colony phase of growth.

According to these ideas, we propose the following
function for $D_b$:
\begin{equation}
D_b=D ({\bar d}_1+{\bar d}_2 {\bar b}) {\bar n},
\label{dif}
\end{equation}
where $D$ depends on the concentration of agar,
which is lower for harder medium. To take
into account the inhomogeneities of the medium, we
introduce a quenched disorder in $D$,
which is written as $D=D_0(1+\xi ({\bf r}'))$, 
$\xi ({\bf r}')$ being a random term
defined on a square lattice. From now on, $D_0$ will 
be referred to as the diffusion parameter.

The first term of Eq.\ (\ref{dif}) describes the usual
diffusion of bacteria in a liquid medium.
The second describes the cooperative
enhanced-movement mechanism promoted by long bacteria.
 This second
mechanism can be modeled by a diffusion coefficient
that depends
on the bacterial concentration.
We multiply both terms by nutrient concentration to
take into account the fact that bacteria
are inactive
in the region where nutrient has been depleted. This
dependence on ${\bar n}$ would not have been necessary
if we had considered a "death" term in the equation
for ${\bar b}$. The coefficients $d_1$ and $d_2$ can adopt
two different values (one of them zero)
depending on the concentration of the long bacteria,
as will be specifed below.

Equations  (\ref{mod1}) with Eqs.\
(\ref{growth})-(\ref{dif}) can
be written in a simpler
form as
\begin{eqnarray}
\frac{\partial b}{\partial t}&=& \nabla \left \{ D \left( d_1
+d_2 b \right ) n \nabla b \right \} +
\frac{nb}{1+\gamma n},
\nonumber \\
\label{model} \\
\frac{\partial n}{\partial t}&=& \nabla^2 n
-\frac{nb}{1+\gamma n},
\nonumber
\end{eqnarray}
with
\begin{eqnarray}
{\bf r}&=&\left( \frac{\theta k^2}{D_n} \right )^{1/4} {\bf r}'
~,~~~
t = k \left ( \theta D_n \right)^{1/2} \tau ~, 
\nonumber \\
n&=&\left (\frac{\theta}{D_n}\right )^{1/2}{\bar n}~,~~~
b=\left (\frac{1}{\theta D_n}\right )^{1/2}{\bar b}~, 
\label{dimen} \\
\gamma& =& \left ( \frac{D_n}{\theta} \right )^{1/2} \gamma'~,~~~
d_1=\left (\frac{1}{\theta D_n}\right )^{1/2}{\bar d}_1~, ~~~
d_2={\bar d}_2.
\nonumber
\end{eqnarray}

At this point, we need to specify how to choose $d_1$
and $d_2$
depending on the population of long bacteria.
In order to do this, we introduce a global 
phenomenological quantity $W(t)$ that
measures the amount of long bacteria.
The evolution of this quantity should have a "creation
term" that represents the
transformation of short bacteria into long ones, and
an "annihilation term" that
represents the opposite transformation (septation).
It seems reasonable to assume that
the creation term is directly dependent on the mean
bacterial concentration,
and inversely dependent on the level of nutrient
($n_0$) and on the diffusion parameter
(adverse conditions, {\it i.e.} $D_0$ and $n_0$ small,
means a faster differentiation process).
With regard to the annihilation term, it can adopt two
possible values
depending on the growth phase.
The simplest equation that includes all these
considerations can be written as
\begin{eqnarray}
\frac{\partial W}{\partial t}&=&\lambda \frac{B}{n_0
D_0}-c_i,
\label{population}
\end{eqnarray}
where $\lambda$ is a constant and $c_i$ can have two
different values ($c_g$ or $c_s$, $c_g>c_s$).
The quantity $B$, defined as
$B=\sum b^2 / \sum b$,
is a measure of the mean concentration of bacteria
inside the colony.
We introduce
the hysteresis previously pointed out by assuming
that there
are two limit values
$W_{MAX}$ and $W_{MIN}$ for which:
\begin{eqnarray}
{\rm when}~~~ W &\ge& W_{MAX} ~~~ {\rm then}~~~ c_i=c_g~,~~~
d_2=D_2~,~~~d_1=0,
\nonumber \\
\label{cond} \\
{\rm when}~~~ W &\le& W_{MIN} ~~~ {\rm then}~~~ c_i=c_s~,~~~
d_2=0~,~~~d_1=1.
\nonumber
\end{eqnarray}

With a suitable choice of parameters $\lambda$,
$c_s$ and $c_g$, 
and by changing only $n_0$ and $D_0$,
we can obtain colonies that
always move with one of the two types of diffusion, or
colonies that periodically
change from one type to the other. This 
will occur if $c_s < \lambda B / (n_0 D_0) < c_g$, which 
will give rise to the ring patterns.
Although the bacterial response has been expressed in terms
of the population of long-bacteria, other possible
kinds of responses admit identical modelization. In this
sense, our model is quite general.

\section{Numerical results}

We have numerically integrated Eqs.\
(\ref{model})
with (\ref{population})-(\ref{cond})
in a square lattice of lateral size $L=600$ using a 4-th order
Runge-Kutta's method
with mesh-size $\Delta x=0.5$ and time step 
$\Delta t=0.005$.
The system was initially prepared by assigning to each
point a nutrient concentration
$n({\bf r},0)=n_0+\eta({\bf r},0)$,
$\eta$ being a uniform random
number in the interval $(-0.1,0.1)$, and a bacterial
concentration $b({\bf r},0)=0$,
except in a small central square
where $b({\bf r},0)=b_0$.
The random term of the diffusion,
$\xi({\bf r})$, takes a different and uncorrelated
value in each box of side $4\Delta x$. The random
values are assumed
to be uniformly distributed in the interval
$(-\epsilon,\epsilon)$.
The box size and the intensity $\epsilon$
do not essentially affect the results.

In all our simulations, we used the parameters
$b_0=0.7$, $\gamma=0.5$, $D_2=30$,
$\epsilon=0.4$,
$\lambda=0.18$, $c_g=2$, $c_s=1.6$, $W_{MAX}=3$
and $W_{MIN}=2$.
We reproduce the different morphologies observed in
experiments by changing the values of the initial
concentration of nutrients $n_0$ and the softness of
the media, related to $D_0$.

\begin{figure}
\begin{center}
\def\epsfsize#1#2{0.50\textwidth}
\leavevmode
\epsffile{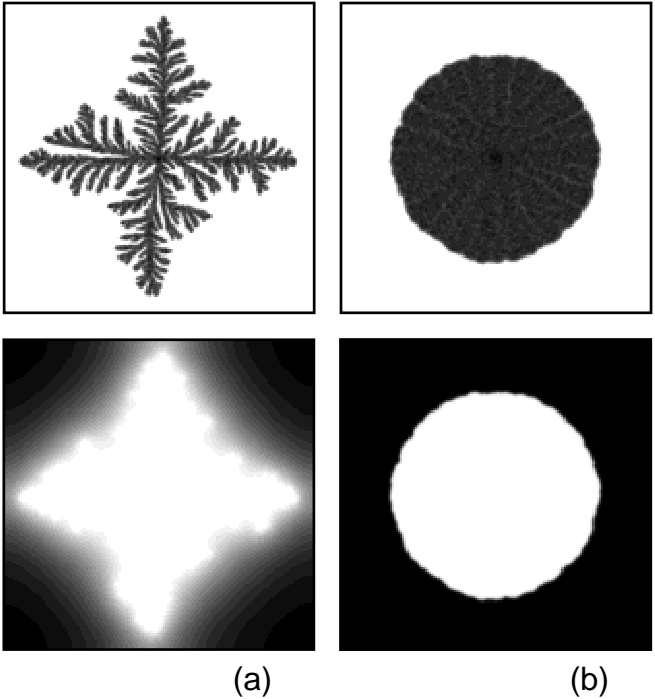}
\end{center}
\caption{
\label{fig2}
Bacterial colonies (top) and nutrient patterns (bottom)
for a fixed value of $D_0=0.01$ and two values
of initial nutrient:
$n_0=1 (a)$ and $n_0=5 (b)$.
They correspond to times $t=2500$ and $75$
respectively.}
\end{figure}
\begin{figure}
\begin{center}
\def\epsfsize#1#2{0.50\textwidth}
\leavevmode
\epsffile{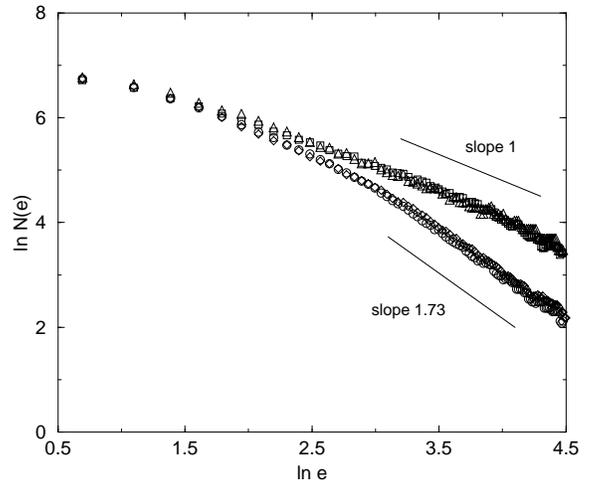}
\end{center}
\caption{
\label{fig3}
Fractal dimensions determined by the box--counting method,
for $D_0=0.01$ and $n_0=1 (\bigcirc),2 (\Diamond),
3 (\Box)$ and $5 (\triangle)$. Two lines of slopes
$1$ and $1.73$ are also plotted for comparison with
experiments.}
\end{figure}

In Figs. \ref{fig2}-\ref{fig3} we present the results obtained for
$D_0=0.005$
and different values of $n_0$.
By increasing $n_0$ we reproduce the crossover between regions $A$ and $B$
of Fig.\ \ref{fig1}, from DLA-like patterns (Fig.\ \ref{fig2}(a)) to
a dense rough structure
similar to that found with an Eden model
(Fig.\ \ref{fig2}(b)).
All of them correspond to a situation in which, due to the small value of
$D_0$, the creation term of Eq.\ (\ref{population}) is
greater than $c_g$, except at the very beginning.
The response $W$ can never decrease below
the value $W_{MIN}$, and therefore the colony will always grow with the
enhanced-movement mechanism. In spite of this cooperative mechanism,
and because of the hardness of the medium, the effective
diffusion $D_0 D_2 b n$ is still small. The growth is
mostly due to
reproduction by feeding on the nutrient.
For low level of nutrient,
{\it i.e.} small $n_0$, the colony growth is limited by
the diffusion of these nutrients.
It develops branches, which are thicker as $n_0$
increases.
The prototype model that reproduces this kind of
structure is the
diffusion-limited aggregation
(DLA) \cite{witten}, which is known to form a fractal
pattern with a fractal dimension of
$d_F=1.71$\cite{Mat,barabasi}.
Experiments performed by Matsushita et al. \cite{Matsus}
in region $A$ of Fig.1 also show a fractal growth with
dimension $d_F=1.73$ \cite{Matsus}.
We have analyzed the fractal nature of the patterns obtained
with our model,
for $D_0=0.005$ and several values of the initial nutrient,
from $n_0=1$ (DLA-like) to $n_0=5$ (rough structure).
We have calculated their fractal dimensions by using the
box-counting method \cite{Mat,barabasi}. In Fig.\ \ref{fig3}
we show, in a log-log plot, the number $N$ of boxes of size
$e$ that contains any part of the pattern, versus the size of
the boxes. The slopes of the lines represent the fractal dimensions.
We observe that the cases that correspond to low nutrient
have a fractal dimension of about $d_F=1.73$, showing good agreement
with experiments. On the other hand, there is
an abrupt  change between these patterns and those
that are not fractal ($d_F=1$). These last cases can be 
analyzed in terms of  the roughness of their interfaces.

\begin{figure}[h]
\begin{center}
\def\epsfsize#1#2{0.50\textwidth}
\leavevmode
\epsffile{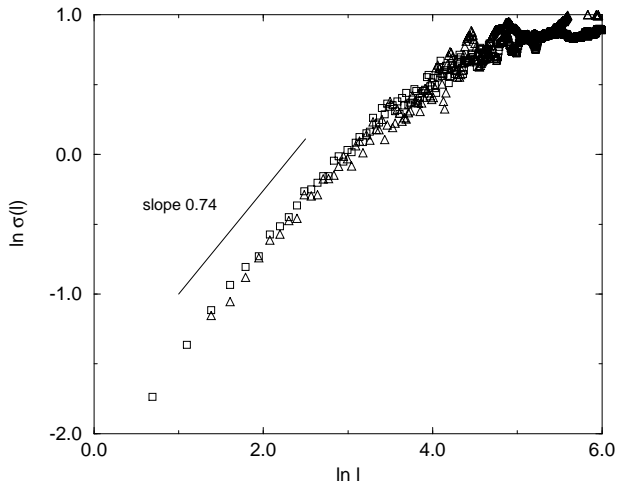}
\end{center}
\caption{
\label{fig4}
Width of the rough interface
as a function of the length of
the interval in which it is measured.
They correspond to the pattern of
Fig.\ \ref{fig2}(b) ($\Box$) and to a case
with the same parameters but with strip
geometry ($\triangle$).
A line of slope $0.74$ is also plotted
for comparison with experimental results.}
\end{figure}

It is well known that Eden structures are not themselves fractal, but
their surfaces exhibit a self-affine scaling \cite{barabasi,krug}. This
implies that, for a long enough time, the width of the rough interface
$\sigma$ scales with an exponent $\alpha$ as a function
of the length of the interface $l$ ($\sigma \sim l^\alpha$).
The roughness exponent for the
Eden model is $\alpha=0.5$.
Vicsek et al. \cite{vicsek} analyzed experimental data corresponding
to the region $B$ of Fig.\ \ref{fig1}. They concluded that
these colony surfaces are self-affine with a roughness exponent
$\alpha=0.74$.
We have checked this point for our dense rough pattern 
(Fig.\ \ref{fig2}(b)) by
measuring the width $\sigma$ for
intervals of interface of length $l$. The results, as a function
of $l$, are presented in Fig.\ \ref{fig4}.
In order to avoid additional effects derived
from the radial growth of the colony, we have also performed 
a complementary
simulation for the same parameters as Fig.\ \ref{fig2}(b)
but with a strip geometry.
To do this, we have used a rectangular lattice of horizontal lateral size
$L_x=600$,
with periodic boundary conditions in $x$ direction, and taken as an initial
condition for bacteria a horizontal line of length $L_x$. The results for this
case are
also plotted in Fig.\ \ref{fig4}. For both circular and strip cases, we observe
analogous behavior to that
observed in experiments \cite{vicsek}. Our results show
a linear region with a slope compatible with the experimental
value $\alpha=0.74$.

\begin{figure}[h]
\begin{center}
\def\epsfsize#1#2{0.50\textwidth}
\leavevmode
\epsffile{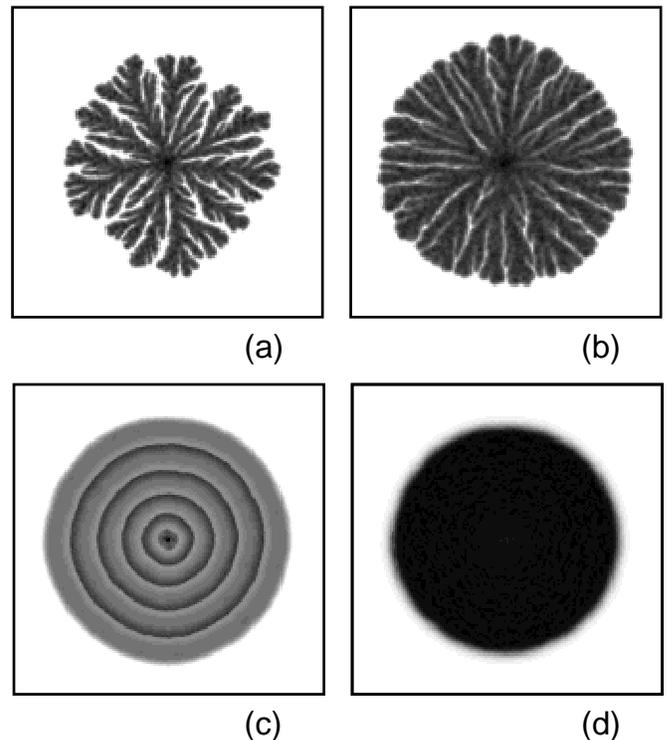}
\end{center}
\caption{
\label{fig5}
Patterns  obtained
for $n_0=1$ with
$D_0=0.05 (a)$ and $0.1 (b)$ and
for $n_0=5$ with $D_0=0.1 (c)$ and $2 (d)$.
They correspond to times $t=500, 300, 50$ and $25$
respectively.}
\end{figure}

With the aim of reproducing other morphologies of Fig.\ \ref{fig1},
we now keep the initial nutrient fixed at the value $n_0=1$ and
increase the diffusion parameter $D_0$.
Results are shown in Fig.\ \ref{fig5}(a)-(b).
We observe a crossover from the 
DLA-like structure (Fig.\ \ref{fig2}(a)) to a dense branching
morphology analogous to that represented in
region $E$ of Fig.\ \ref{fig1}.

In Fig.\ \ref{fig5}(c)-(d), we present two snapshots obtained for
a fixed value of the initial nutrient $n_0=5$.
They show how different kinds of patterns are obtained
when $D_0$ is increased:
from the dense rough structure (Fig.\ \ref{fig2}(b)),
to concentric rings (Fig.\ \ref{fig5}(c)) and homogeneous disk
(Fig.\ \ref{fig5}(d)). They correspond to the regions $B, C$ and $D$
respectively.
Homogeneous disks are obtained when $D_0$ and $n_0$ are so high
that the creation term of Eq.\ (\ref{population}) is always smaller than
$c_s$. This means that the value $W_{MAX}$, above which the
enhanced-movement mechanism begins, is never reached, and bacteria move
with the usual diffusion coefficient $D_0 n$.

\begin{figure}[h]
\begin{center}
\def\epsfsize#1#2{0.50\textwidth}
\leavevmode
\epsffile{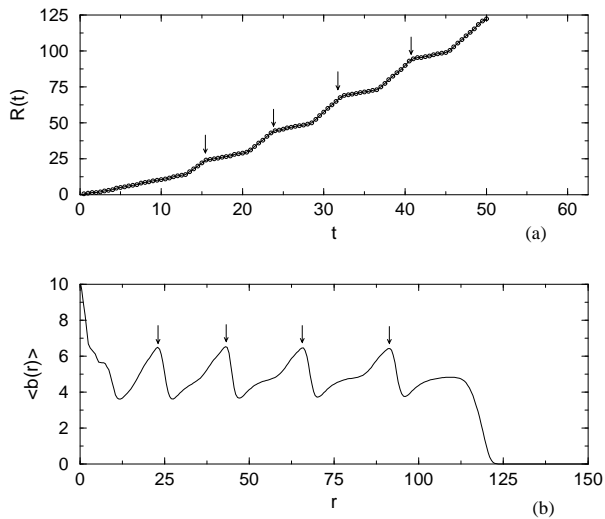}
\end{center}
\caption{
\label{fig6}
$(a)$ Time evolution of the colony radius, for parameters
$D_0=0.1$ and $n_0=5$;
$(b)$ Radial density profile corresponding to time $t=50$
(pattern of Fig.\ \ref{fig5}(c)).
The positions where each consolidation phase started
are pointed out.}
\end{figure}

Ring patterns correspond to a narrow region of parameters
$D_0$ and $n_0$ for which the creation term of Eq.\ (\ref{population})
takes a value between $c_s$ and $c_g$. As explained in Section II,
this leads to dynamics in which
bacteria move alternatively by usual diffusion $D_0 n$ (consolidation phase)
or by the enhanced-movement mechanism $D_0 D_2 b n$ (migration phase).
The two phases are clearly manifested in
Fig. \ref{fig6}(a), where we represent the radius of
the colony as a function of time.
The pattern of concentric rings is a consequence of this dynamic behavior.
In Fig.\ref{fig6}(b) we plot the radial density profile, circularly-averaged,
corresponding to the ring pattern shown in Fig.\ \ref{fig5}(c). The maxima
are formed in the positions where a consolidation phase began. To
illustrate this point, we have pointed out in Fig.\ \ref{fig6} the positions
corresponding to the colony radius at the beginning of each consolidation phase.

Numerically, our model also reproduces the experimentally observed
robustness of the growth-plus-consolidation period, which is barely 
dependent on changes in either nutrient or agar concentrations
over a wide range. For high enough $n_0$, the value of the
global quantity $B$ approaches $n_0$. In this limit, as
can be derived from Eq.\ (\ref{population}) the period 
is given by
\begin{eqnarray}
T=D_0 (W_{MAX}-W_{MIN})\left (\frac{1}{\lambda-D_0 c_s}-
\frac{1}{\lambda-D_0 c_g}\right),
\label{period}
\end{eqnarray}
which does not depend on $n_0$.
Moreover, as a function of $D_0$, the period
also maintains a rather constant value within a certain range 
(determined by parameters $\lambda$, $c_g$ and $c_s$) to 
increase sharply in the boundaries of 
the ring patterns region ($D_0 \longrightarrow {\lambda / c_s}$,
$D_0 \longrightarrow {\lambda / c_g}$). For equal period, the width 
of the rings increases with $n_0$.

\section{Conclusions}

We have proposed  a reaction-diffusion model for the study of bacterial
colony growth on agar plates, which consists of
two coupled equations for nutrient and bacterial concentrations.
The most important feature, which introduces differences from previous
models, is the fact that here we consider two mechanisms for the bacterial
movement: the random swimming in a liquid medium, and a cooperative
enhanced movement developed by bacteria when the growth conditions
are adverse. The two mechanisms are introduced in our model by means
of a diffusion term with two different expressions which
depend on the bacterial response to the environmental conditions.
This response is modeled as a global variable that
presents hysteresis depending on the conditions of the medium.
The inhomogeneities of the agar plate have been taken into account as
a quenched disorder in the diffusion parameter.

We have shown that, simply by changing the parameters related to the
hardness of the medium and the initial nutrient, our model reproduces all
the patterns obtained experimentally with the bacterium {\it Bacillus subtilis}:
DLA-like, dense-rough disk, DBM-like, ring patterns and homogeneous disk.
We have calculated the fractal dimension of the DLA-like structures and the
roughness exponent of the rough disk surface, obtaining results in good 
agreement with experiments. The ring patterns have been obtained for
intermediate values of agar and high nutrient. In this region,
the bacterial response presents hysteresis and
the two mechanisms of motion work alternatively, leading to cycles
of migration and consolidation phases. The duration of these cycles
is roughly constant for different
values of nutrient and agar concentration over a wide range.
This periodical dynamics generates
patterns of concentric rings.

In summary, the model proposed satisfactorily reproduces
the whole experimental morphological diagram. It represents a first
attempt at describing the response of bacteria to adverse growth conditions
and, in certain conditions, their ability to improve their motility.
Further refinements could be made. 
The bacterial response, here described
as a global variable $W(t)$, could be considered in a more realistic way 
by introducing a coupling term  in a local version of
Eq.\ (\ref{population}) for a field $W({\bf r},t)$. 
However, preliminar studies with such a model \cite{lacasta} shows 
the same essential features previously described.

\acknowledgements

We thank I. R\`afols and J.M. Sancho for helpful
discussions.
This research was supported by
Direcci\'on General de Investigaci\'on Cient\'{\i}fica
y T\'ecnica (Spain) (PB96-0241-02),
by Comissionat per Universitats i Recerca de la
Generalitat de Catalunya (SGR97-439) and by Universitat
Polit\`ecnica de Catalunya (PR-9608).
We also acknowledge computing support from
Fundaci\'o Catalana per a la Recerca and
Centre Catal\`a de Computaci\'o i Comunicacions.

\end{document}